\documentclass[aps,twocolumn,prl,superscriptaddress]{revtex4}

\usepackage{amssymb,amsmath}

\begin{document}

\title{Maximally Non-Local and Monogamous Quantum Correlations}

\author{Jonathan Barrett}
\email[]{jbarrett@perimeterinstitute.ca}
\affiliation{Perimeter Institute for
  Theoretical Physics, 31~Caroline Street N, Waterloo, Ontario N2L 2Y5,
  Canada}
\author{Adrian Kent}
\email[]{a.p.a.kent@damtp.cam.ac.uk}
\affiliation{Centre for Quantum Computation, DAMTP, Centre for Mathematical
  Sciences, University of Cambridge, Wilberforce Road, Cambridge CB3 0WA,
  U.K.}
\author{Stefano Pironio}
\email[]{Stefano.Pironio@icfo.es}
\affiliation{ICFO-Institut de Ci\`{e}ncies Fot\`{o}niques, Mediterranean
  Technology Park, 08860 Castelldefels (Barcelona), Spain\\ Institute for
  Quantum Information, California Institute of Technology, Pasadena, CA 91125,
  U.S.A.}

\begin{abstract}
  We introduce a version of the chained Bell inequality for an arbitrary
  number of measurement outcomes, and use it to give a simple proof that the
  maximally entangled state of two $d$ dimensional quantum systems has no
  local component.  That is, if we write its quantum correlations as a mixture
  of local correlations and general (not necessarily quantum) correlations,
  the coefficient of the local correlations must be zero. This suggests an
  experimental programme to obtain as good an upper bound as possible on the
  fraction of local states, and provides a lower bound on the amount of
  classical communication needed to simulate a maximally entangled state in
  $d\times d$ dimensions.  We also prove that the quantum correlations
  violating the inequality are monogamous among non-signalling correlations,
  and hence can be used for quantum key distribution secure against post-quantum
  (but non-signalling) eavesdroppers.
\end{abstract}

\date{May 19, 2006}
\maketitle

Quantum theory predicts that measurements on separated entangled systems will produce
outcome correlations that, in Bell's terminology \cite{bellbook}, are not {\it
  locally causal} or, in what has become standard terminology, are {\it
  non-local}.  In particular, if a Bell inequality is violated then one cannot
consistently assume that the outcomes of measurements on each system are
predetermined and independent of the measurements carried out on the
other system(s).  Violation of Bell inequalities has been confirmed in numerous
experiments \cite{exp}.

Violation of Bell inequalities not only tells us something fundamental about
Nature, but also has practical applications.
For example, the nonlocality of quantum correlations
allows communication complexity problems to be solved using an amount of
communication that is smaller than is possible classically \cite{commcompl}.
Barrett-Hardy-Kent (BHK) recently showed that testing particular nonlocal quantum correlations
allows two parties to distribute a secret key securely,
in such a way that the security is guaranteed by the no-signalling principle
alone \cite{qkd} (i.e. without relying on the validity of quantum theory).

In this paper, we extend the chained Bell inequality to an inequality with an
arbitrary number of measurement outcomes, which can thus be applied to states
in arbitrary dimensions. In the limit of a large number of measurement
settings, quantum mechanics predicts correlations for a maximally entangled
bipartite state that resemble those of the tripartite
Greenberger-Horne-Zeilinger (GHZ) state in that, with probability tending to
one, the predictions of any local hidden variable model contradict those of
quantum mechanics for at least one pair of measurements. We use this to give a
constructive proof of a result originally due to Elitzur, Popescu and Rohrlich
(EPR) \cite{epr}: if the quantum correlations of a maximally entangled state
of two qubits are written as a convex combination of local and nonlocal
correlations, then the local fraction must be zero.  Our proof extends EPR's
result to maximally entangled states in any dimension and also removes the
need for a technical assumption required for EPR's original proof
\cite{footnote1}.  Moreover, because our proof is constructive, it motivates
an experimental programme to establish the best possible upper bound on the
fraction of local states in a maximally entangled state.
More generally, our proof method works, and motivates experimental
tests, for any example of GHZ-type correlations.

Next, we give a rigorous proof of the \emph{monogamy} of the
correlations obtained from a $d\times d$-dimensional maximally entangled
state.  Here, monogamy means that a third
party cannot get any information about the measurement outcomes,
so long as a no-signalling
condition holds, even if quantum theory is incorrect.
This property has a particular significance in the
context of secret key distribution, and our results here generalise
those of BHK \cite{qkd}, which proved monogamy in the $2 \times 2$-dimensional
case in order to demonstrate the security of BHK's scheme against general
non-signalling eavesdroppers.

Finally, as a corollary, we derive a lower bound
on the classical communication needed to simulate measurements on a maximally
entangled state in $d\times d$ dimensions.

\paragraph{Some chained Bell inequalities.}
Consider a standard Bell-type experiment: two parties, Alice and
Bob, share a joint system in an entangled quantum state, and perform
measurements on their local subsystems.  Each party
may choose one out of $N$ different measurements, and each
measurement $A_k$ of Alice and $B_l$ of Bob ($k,l=1,\ldots,N$) may
have $d$ possible outcomes: $A_k,B_l=0,\ldots,d-1$.
Quantum theory predicts the joint probabilities $P^{\scriptstyle QM}(A_k=a,B_l=b)$ that
Alice's and Bob's measurements, $A_k$ and $B_l$, have
respective outcomes $a$ and $b$.
The correlations predicted by quantum theory are {\it local}
if and only if they can be written in the form
\begin{equation}\nonumber
P^{\scriptstyle QM}(A_k=a,B_l=b)=
\sum_{\lambda} q_{\lambda} P_{\lambda}(A_k=a)\times P_{\lambda}(B_l=b) \, ,
\end{equation}
with $0\leq q_{\lambda}\leq 1$ and $\sum_{\lambda} q_{\lambda}=1$.
Without loss of generality, we can assume that the terms
$P_{\lambda}(A_k=a)$ and $P_{\lambda}(B_l=b)$ are deterministic, that is, that
they only take the values $0$ or $1$ \cite{ww}. The correlations are thus
local if and only if they can be reproduced by a mixture of hidden states
assigning definite values to each measurement. Violation of a Bell
inequality implies that the correlations cannot be written in this form.

In general one can write the correlations as
\begin{multline}\label{mixture}
P^{\scriptstyle QM}(A_k=a,B_l=b)=p\,P^{\scriptstyle L}(A_k=a,B_l=b)\\
+(1-p)\,P^{\scriptstyle NL}(A_k=a,B_l=b)\,,
\end{multline}
where $P^{\scriptstyle L}(A_k=a,B_l=b)$ and $P^{\scriptstyle NL}(A_k=a,B_l=b)$
represent local and non-local joint distributions, respectively, and $p$ and
$(1-p)$ their corresponding weights in the mixture, with $0 \leq p \leq 1$.
Suppose now that we have a set of quantum correlations which satisfy some
relation with certainty, and we can show that any local model must fail to
satisfy the relation at least some of the time.  It follows that
the correlations cannot be written in the form of
Eq.~\eqref{mixture} except with $p=0$.
(Note that this is true even if the
term $P^{\scriptstyle NL}(A_k=a,B_l=b)$ is allowed to describe signalling
correlations.)
A particularly well-known example of such correlations was that produced by
GHZ \cite{ghz} and simplified by Mermin \cite{mermin};
hence we refer to correlations with these properties as
GHZ-type correlations.
Bipartite examples have been given by Heywood-Redhead \cite{heywoodredhead} and
Cabello \cite{cabello}.
A set of quantum correlations is of GHZ-type if and only if
it has an associated Bell inequality which is violated right up to the
algebraic limit of the expression.

We now derive a Bell inequality and show that in the limit of a large number
of measurement settings, the quantum correlations from a maximally entangled
state violate the inequality up to the algebraic limit. Thus the quantum
correlations tend to GHZ-type as the number of
measurement settings becomes large.  Hence, we show,
maximally entangled states in any dimension have zero local component.

Consider first the case where Alice and Bob each have a choice between
two different measurements ($N=2$). Local correlations satisfy the following inequality
\cite{cglmp},
\begin{multline}\label{cglmp}
I_2=\langle[A_1-B_1]\rangle+\langle[B_1-A_2]\rangle+\langle[A_2-B_2]\rangle\\
+\langle[B_2-A_1-1]\rangle \geq d-1\, ,
\end{multline}
where
$\langle X \rangle= \sum_{i=1}^{d-1} i P\,(X=i)$
is the average value of the random variable $X\in\{0,\ldots,d-1\}$
and $[X]$ denotes $X$ modulo $d$.
This follows from the identity
\begin{equation*}
[A_1 - B_1 + B_1 - A_2 + A_2 - B_2 + B_2 - A_1 - 1]=d-1\,,
\end{equation*}
and the relation $[X]+[Y]\geq [X+Y]$. When $d=2$, it corresponds to
the CHSH inequality \cite{chsh}.

We can extend the above
inequality to an arbitrary number $N$ of measurement choices:
\begin{multline}\label{chained}
I_N=\langle[A_1-B_1]\rangle+\langle[B_1-A_2]\rangle+\langle[A_2-B_2]+\cdots\\
\cdots+\langle[A_N-B_N]\rangle+\langle[B_N-A_1-1]\rangle\geq d-1\, .
\end{multline}
This extended inequality can be viewed as a chained version of
inequality \eqref{cglmp}, and follows
by a similar argument. It is equivalent,
when $d=2$, to the chained inequality introduced by
Pearle \cite{pearle} and Braunstein and Caves \cite{bc}.
The relation between the Pearle-Braunstein-Caves
inequality and GHZ-type correlations was
first pointed out by Hardy \cite{hardy}.

We now show that if Alice and Bob share the maximally entangled state
\begin{equation}\label{mes}
|\psi_d\rangle=1/\sqrt{d}\,\sum_{q=0}^{d-1}|q\rangle_A|q\rangle_B \, ,
\end{equation}
there exist measurement settings such that for large $N$,
$I_N({\scriptstyle
  QM})$ tends to zero. The maximally entangled state $|\psi_d\rangle$ has the
property that if Alice measures an observable with eigenvectors $|r\rangle$
($r=0,\ldots,d-1$) and Bob measures the observable with complex conjugate
eigenvectors $|r\rangle^*$, they get perfectly correlated outcomes.  We define
the eigenvectors characterizing Alice's measurement $A_k$ as
\begin{equation}\label{ka}
|r\rangle_{A_k}=\frac{1}{\sqrt{d}}\sum_{q=0}^{d-1}
\exp\left[\frac{2\pi i}{d}q(r-\alpha_k)\right]|q\rangle_A\,,
\end{equation}
and those characterizing Bob's measurement $B_l$ as
\begin{equation}\label{kb}
|r\rangle_{B_l}=\frac{1}{\sqrt{d}}\sum_{q=0}^{d-1}
\exp\left[-\frac{2\pi i}{d}q(r-\beta_l)\right]|q\rangle_B\,,
\end{equation}
where $\alpha_k=(k-1/2)/N$ and $\beta_l=l/N$.

A straightforward calculation shows that each term in \eqref{chained} is equal
to $\gamma/N^2+O(1/N^3)$, where $\gamma=\pi^2/(4d^2)\sum_{j=1}^{d-1}
j/\sin^2(\pi j/d)$.  As there are $2N$ such terms in the inequality
\eqref{chained}, it follows that
\begin{equation}\label{qmin}
I_N({\scriptstyle QM})= 2\gamma/N+O(1/N^2),
\end{equation}
which can be made arbitrarily small for sufficiently large $N$.

For any particular finite value of $N$, the quantity $I_N({\scriptstyle QM})$
implies an upper bound on the fraction of local states in any model that
reproduces the correlations, i.e., an upper bound on the $p$ of
Eq.~\eqref{mixture}. This is because although the term $P^{\scriptstyle NL}(A_k=a,B_l=b)$
can violate the Bell inequality \eqref{chained}, it must satisfy
$I_N({\scriptstyle NL})\geq 0$, since each term in \eqref{chained} is a
positive quantity. Thus we can write $I_N({\scriptstyle
  QM})=p\,I_N({\scriptstyle L})+(1-p)\,I_N({\scriptstyle NL})$, and, since
$I_N({\scriptstyle L})\geq d-1$ and $I_N({\scriptstyle NL})\geq 0$,
$I_N({\scriptstyle QM})\geq p\,(d-1)$, or
\begin{equation}\label{bound}
p \leq \frac{I_N({\scriptstyle QM})}{d-1}  \, .
\end{equation}
Of course, in a real experiment, the state prepared
will not be precisely (\ref{mes}), the measurements will not be
precisely defined by projections onto the vectors (\ref{ka}) and (\ref{kb}),
and so on, and thus the experimentally determined
value, $I_N ({\scriptstyle EXP})$, will generally
be greater than $I_N({\scriptstyle QM})$.  Nonetheless,
given a value for $I_N ({\scriptstyle EXP})$, we can
obtain a bound of the form
\begin{equation}\label{expbound}
p \leq \frac{I_N({\scriptstyle EXP})}{d-1}  \, .
\end{equation}
Hence we propose an experimental challenge: to obtain
the lowest possible bound on $p$ (for any $d$ and $N$)
for bipartite maximally entangled states.

Our proof extends to any example of
GHZ-type correlations; thus another natural challenge
is to obtain the lowest possible bound on $p$ for any
set of quantum correlations on any bipartite or multipartite entangled state.

\paragraph{Indeterminacy of the measurement outcomes and Monogamy.}
The correlations that we just introduced have a particular significance in the
context of key distribution.  BHK showed \cite{qkd} how non-local correlations
can be used as the basis of a key distribution scheme that is secure against
non-signalling eavesdroppers.  Non-local correlations can also be used to give
at least partial security against non-signalling eavesdroppers in more
practical QKD schemes \cite{bhkunpub,acin}.  In these discussions, it is not
assumed that such eavesdroppers are constrained by the laws of quantum
mechanics, but it is assumed that they can only prepare systems in states
whose correlations are {\it non-signalling}, in the following sense.  Suppose
that Alice and Bob share a bipartite system characterized by correlations
$P(A_k=a, B_l=b)$.  The correlations are non-signalling if they satisfy
\begin{equation}\label{nosig}
\sum_a P(A_k=a, B_l=b) = \sum_{a'} P(A_{k'}=a', B_l=b) \, ,
\end{equation}
for all $k,k',l,b$, and a similar set of conditions obtained by summing
over Bob's input. These no-signalling conditions ensure that the marginal
distributions $P(A_k=a)$ and $P(B_l=b)$ are well defined quantities. The
definition of no-signalling can be extended to more than two parties,
by requiring a condition similar to \eqref{nosig} for every
possible grouping of the parties into two subsets.

In the protocols of Refs.~\cite{qkd} and \cite{bhkunpub,acin},
maximally entangled states are prepared by a
source, which is situated between Alice and Bob and assumed to be under
control of the eavesdropper Eve. On each pair of particles, Alice and Bob
perform measurements $A_k$ and $B_l$, chosen independently and randomly, and
use the corresponding measurement outcomes to establish a shared secret key.

If the correlations used to distribute the key admit a model with
fractions $p$ and $1-p$ of local and non-signalling non-local states respectively,
then for a fraction $p$ of pairs, Eve could prepare a deterministic local
state that would give her complete information about Alice's and Bob's
measurement outcomes. This strategy is clearly not significantly useful to Eve if
the quantum correlations imply $p\approx 0$, which is the case in BHK's protocol \cite{qkd},
as noted above.

However, a zero fraction of local states does not necessarily imply that Eve's information
is zero, as it does not exclude the possibility that she could prepare
a non-local state that has definite values for a non-empty proper subset of the
measurement inputs. For instance, in the case of Cabello's example \cite{cabello},
there is a model reproducing the quantum correlations with
the following properties: i) every hidden state is non-signalling, ii) for
every hidden state, at least one measurement has a predetermined outcome, iii)
each measurement has a predetermined outcome for at least some finite fraction
of the hidden states. A non-signalling Eve exploiting this model could obtain
some knowledge of Alice's and Bob's measurement outcomes. Similar remarks
apply to the GHZ and Mermin examples.

The correlations defined by \eqref{mes}, \eqref{ka} and \eqref{kb},
however, are stronger in the sense that
any non-signalling model that reproduces the correlations has the following property:
for every hidden state, the outcomes of any measurement are completely undetermined and
occur all with the same probability $1/d$. This property is implied by the following theorem.\\
\textbf{Theorem.}{  \it Any no-signalling distribution for which ${I}_N\leq I_N^*$
satisfies}
\begin{equation}\nonumber
P(A_k=a)\leq \frac{1}{d}+\frac{d}{4} I_N^* \quad \mbox{and} \quad P(B_l=b)
\leq \frac{1}{d}+\frac{d}{4} I_N^*
\end{equation}
{\it for all measurements $A_k$ and $B_l$ and for all outcomes $a$ and $b$.}

In the limit $I_N^* \rightarrow 0$, Eve cannot therefore gain any knowledge about
Alice and Bob's measurement outcomes. In other words, any tripartite
distribution describing the joint systems of Alice, Bob and Eve is of the form
$P(A_k=a, B_l=b)\times P(E_m=e)$. We say that Alice's and Bob's correlations
are {\it monogamous}, in obvious analogy with the familiar monogamy of
entanglement.

\emph{Proof of the Theorem.}  Suppose that $P(A_k=a)>1/d+d/4\,I^*_N$ for some
measurement $A_k$ and outcome $a$. We will then show that $I_N>I_N^*$.
(The same argument applies if we suppose that $P(B_l=b)>1/d+d/4\,I^*_N$ for
some measurement $B_l$ and outcome $b$.)

Defining $A_{N+1}=A_1+1$ (modulo $d$), we can write
\begin{equation}\begin{split}
I_N&=\sum_{j=1}^N ( \langle[A_j-B_j]\rangle+\langle[B_j-A_{j+1}]\rangle ) \\
&\geq 2N-\sum_{j=1}^N ( P(A_j=B_j)+P(A_{j+1}=B_j) ) \,,
\end{split}\end{equation}
since $\langle[X]\rangle= \sum_{i=1}^{d-1} i \, P([X]=i) \geq 1-P([X]=0)$. Now
\begin{equation*}\begin{split}
P(A_i=B_j)&=\sum_{r=0}^{d-1} P(A_i=r, B_j=r)\\
&\leq \min(P(A_i=q),P(B_j=q))\\
&\quad + \min(1-P(A_i=q),1-P(B_j=q))\\
&=1-|P(A_i=q)-P(B_j=q)|\,,
\end{split}\end{equation*}
for any $q\in\{0,\ldots,d-1\}$. Using this expression in the above inequality
for $I_N$ and defining $N$ arbitrary different values $q_j$, we get
\begin{equation}\begin{split}\label{int}
I_N&\geq\sum_{j=1}^N ( |P(A_j=q_j)-P(B_j=q_j)|\\
&\qquad+|P(A_{j+1}=q_j)-P(B_j=q_j)| ) \\
&\geq \sum_{j=1}^N|P(A_j=q_j)-P(A_{j+1}=q_j)|\,,
\end{split}\end{equation}
where the second line follows from the triangle inequality. The hypothesis
$P(A_k=a)> 1/d+d/4\,I_N^*$ implies that
$$|P(A_k=a')-P(A_k=a'+1)|>I_N^*$$
for some $a'$.  If we define
the values $q_j$ used in \eqref{int} by $q_1=\ldots=q_{k-1}=a'$ and
$q_k=\ldots=q_{N}=a'+1$ (modulo $d$), we obtain the inequality
\begin{equation}\begin{split}
I_N&\geq|P(A_1=a')-P(A_k=a')\\
&\qquad+P(A_{k}=a'+1)-P(A_{N+1}=a'+1)|\\
&= |P(A_k=a')-P(A_{k}=a'+1)|\\
&>I_N^* \, ,
\end{split}\end{equation}
since $A_{N+1}=A_1+1$ (modulo $d$), by definition. \hfill$\square$

\paragraph{Classical simulation of the correlations.}
Finally, we note that the bound \eqref{bound} can be interpreted, in the
spirit of \cite{pironio}, as a bound on the average communication necessary to
simulate classically non-local correlations. The fact that every hidden state
must be non-local in any model that reproduces measurements on a maximally
entangled state implies that in any run of a communication-assisted classical
protocol some communication must be exchanged between the parties.  Since the
minimal amount of communication that the parties can exchange in any such run
is one bit, it follows that at least one bit of communication is necessary to
simulate these correlations. For qubits, i.e, when $d=2$, one bit is known to
be sufficient \cite{tb}. For other values of $d$, it is possible to extend our
analysis by letting the range of measuring settings go from 1 to $(d-1)\,N$,
while keeping the measurements as defined in \eqref{ka} and \eqref{kb}. Using
inequalities similar to \eqref{chained}, one can then show that at least
$\log_2 d$ bits of communication are necessary. This bound is not optimal
asymptotically since for large $d$ a bound of $O(d)$ bits is known
\cite{brassard}. It may, however, be useful for small $d$.

\paragraph{Summary and conclusions.}
The quantum correlations introduced imply that maximally entangled quantum
states in arbitrary dimensions have zero local component. They motivate
experiments that could bound the weight of the local component as close
to zero as possible.   In a non-signalling context, the correlations are
also provably monogamous, which gives them an immediate application in key distribution.
It would be interesting to characterise
the sets of quantum correlations that are monogamous in this sense;
in particular, it would be interesting to know if there are
monogamous quantum correlations that can be obtained in an experiment with
only a finite number of measurement settings at each site, rather than
in the limit in which the number of settings tends to infinity.

\paragraph{Acknowledgments.} AK thanks Jonathan Oppenheim for helpful conversations.  SP
acknowledges support by the David and Alice Van Buuren fellowship of the
Belgian American Educational Foundation, by the National Science Foundation
under Grant No.~EIA-0086038, and by the European Commission under the
Integrated Project Qubit Applications (QAP) funded by the IST directorate as
Contract Number 015848.


\begin{thebibliography}{21}
\bibitem[{foo()}]{footnote1}
\bibinfo{note}{The assumption made in Ref.~\cite{epr} is that the correlation
  function $E_L(\theta)$ introduced in the paper has an expansion in $\theta$
  for small, positive $\theta$}.
\end{thebibliography}
\end{document}